\documentclass[a4paper,11pt,oneside,fleqn]{article}
   \usepackage[english]{babel}
   \usepackage[utf8]{inputenc}        
   \usepackage[T1]{fontenc}

   \usepackage[usenames,dvipsnames]{color}
   \usepackage{mathpazo}
   \usepackage[scaled=.95]{helvet}
   \usepackage{courier}
   \usepackage[vcentermath]{youngtab}
   \usepackage{amsmath}
   \usepackage{amssymb}
   \usepackage{graphicx}              
   \usepackage{multicol,caption}
   \newenvironment{Figure}
     {\par\medskip\noindent\minipage{\linewidth}}
     {\endminipage\par\medskip}
   \usepackage{url}
   \usepackage{color}                 
\usepackage[table]{xcolor}
   \usepackage{setspace}              
   \usepackage{bm}                    
   \usepackage{lineno}
   \usepackage{rotating}
   \usepackage{tikz} 
   \usetikzlibrary{trees}
   \usetikzlibrary{calc}
   \usetikzlibrary{shapes.geometric}
   \usetikzlibrary{decorations.markings}
   \usetikzlibrary{decorations.text}
   \usetikzlibrary{positioning,shapes,shadows,matrix,arrows}
   \usetikzlibrary{decorations.pathmorphing}
   \usepackage[a4paper,pdftex,top=1cm,bottom=2cm,left=1.0cm]{geometry}
   \usepackage[colorlinks=true,linkcolor=blue,citecolor=blue,urlcolor=blue]{hyperref}

   \newcommand{\beq}{\begin{equation}}
   \newcommand{\eeq}{\end{equation}}

\begin{document}


\pagestyle{plain}
\pagenumbering{arabic}


\begin{flushleft}
{\Large\bf Spectroscopic version of the Aharonov-Bohm effect}
\vspace{0.3cm}
\end{flushleft}

\begin{minipage}{0.85\linewidth}
\begin{spacing}{0.85}
{Caio Laganá}\\[0.1cm]
{\it\footnotesize Nuclear Physics Department, University of Sao Paulo, Brazil}
\vskip 0.7cm
{\small\bf Abstract}\\[0.1cm]
{
\small An experiment is proposed in which the Aharonov-Bohm effect can be veryfied through a spectroscopic measurement. The apparatus consists of gaseous hydrochloric acid (HCl) immersed in the vector potential ${\bf A}$ present in the center of a toroidal coil, where ${\bf B}=0$. Changes due to ${\bf A}$ in the absorption spectrum of the gas are investigated.\\

}
{\footnotesize }
\end{spacing}
\end{minipage}

\singlespacing

\begin{multicols}{2}
	
The Aharonov-Bohm effect (AB) was proposed \cite{ab} and has most frequently been discussed in the literature in the context of non-simply-connected wave functions thread by magnetic fluxes, both in the scattering \cite{ab-scat,ab-toro,ab-var} and bound-state \cite{peshkin,wod,ab-bound-exp} versions of the effect. This topological configuration is mandatory for the scattering setups, since the interference pattern depends on the magnetic flux enclosed by the branches of a splitted electron beam. In the bound-state scenario, however, the observed quantities are energy level differences of a confined system, and the requirement of non-simply-connected wave functions seems to be no longer a requisite. Yet, measurements on metal rings (thus with non-simply-connected wave functions) have been by far the most used devices to study the bound-state AB effect \cite{ab-bound-exp,ring1,ring2,ring3}, and the question of whether the thread-flux topology is an unavoidable requirement has not been given an experimental answer. In this paper, we propose a simple experimental setup to approach this issue.

Besides contributing to the vast literature on the AB effect with a newly setup capable of probing the aforementioned topological issue, the usage of simply-connected wave functions permits the investigation of local aspects of the semiclassical minimal coupling, such as the relative orientation of {\bf A} and the canonical momentum, a feature not possible in setups involving integrals over paths.

\begin{Figure}
 \centering
 \includegraphics[width=\linewidth]{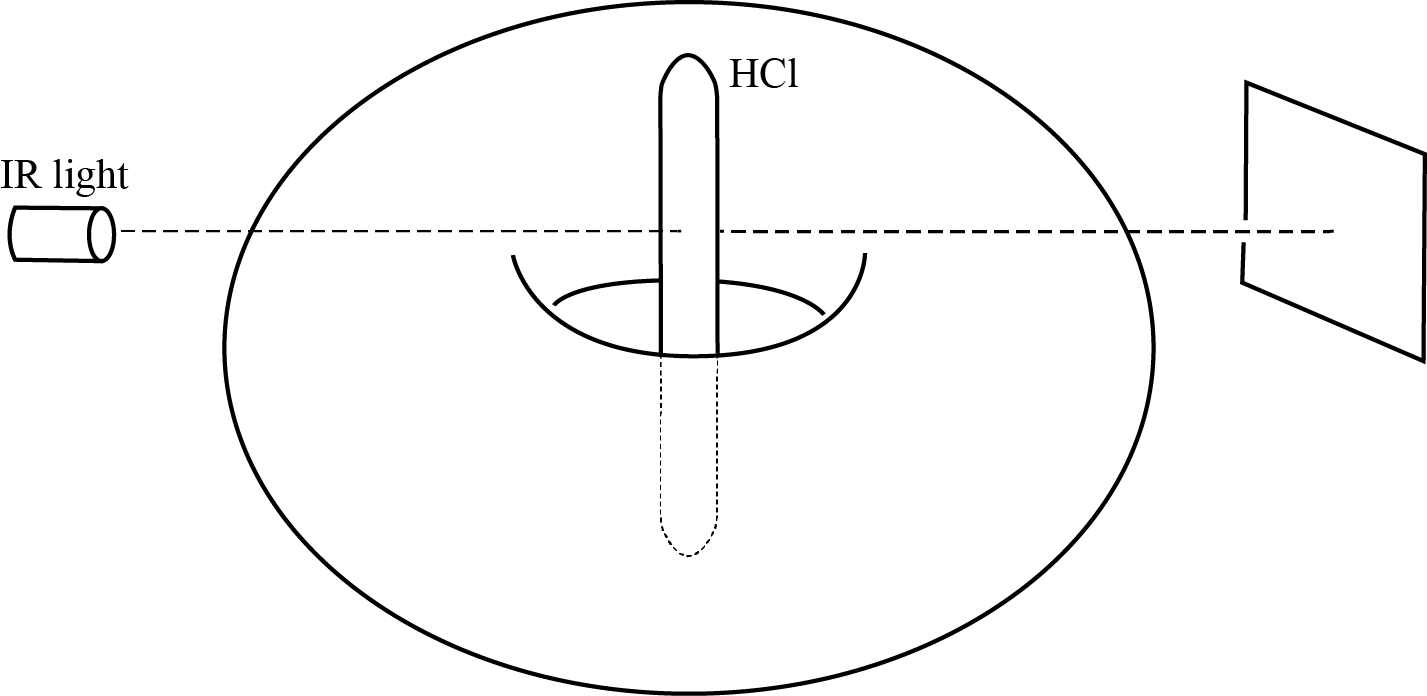}
 \captionof{figure}{\small Proposed setup to attain the AB effect with a simply-connected wave function: measurement of the infrared absorption spectrum of gaseous hydrochloric acid over a toroidal coil.}\label{fig}
\end{Figure}

The vibrational spectrum of HCl molecules is well described by the one-dimensional quantum harmonic oscillator
\begin{equation}\label{H}
  \mathcal{H} = \frac{1}{2\mu}\left[ p^2  + (\mu\omega_0 x)^2\right]
\end{equation}
where $\mu=\tfrac{m_\text{H}m_\text{Cl}}{m_\text{H}+m_\text{Cl}}$ and $\omega_0$ is the characteristic frequency of vibration around equilibrium. $p$ and $x$ represents the relative momentum and position between hydrogen and chlorine. Within each molecule, the vector potential is constant and given by
\beq
  {\bf A}=\frac{\phi}{L}{\bf\hat z}
\eeq
where $\phi$ is the magnetic flux inside the torus and $L$ the perimiter of the curve $\gamma$ such that $\oint_\gamma {\bf A}\cdot d{\bf l}=\phi$. The direction ${\bf\hat z}$ was fixed by the Coulomb gauge. In the presence of ${\bf A}$, Hamiltonian \eqref{H} becomes
\beq\label{eq:H}
   \mathcal{H}=\frac{1}{2\mu}\left[ p^2-2e\frac{\phi}{L}p\cos(\theta) + (\mu\omega_0 x)^2 \right]
\eeq
with $\theta$ the angle between the molecule axis and ${\bf A}$. The constant term $\frac{e^2{\bf A}^2}{2\mu}$ was neglected. Notice that only the component of ${\bf A}$ parallel to the molecule axis was used in the minimal coupling,
\begin{equation}\label{minimal}
   p\to p-eA_0\cos(\theta).
\end{equation}


The matricial representation of \eqref{eq:H} in the usual basis $\{|n\rangle\}$ reads
\begin{align}
   \mathcal{H}&=\hbar\omega_0(a^\dagger a+\frac{1}{2}) -ie\frac{\phi}{L}c\sqrt{\frac{\hbar\omega_0}{2\mu c^2}}\cos(\theta)(a^\dagger-a)\nonumber\\
   &=
\begin{pmatrix}
\frac{1}{2}\hbar\omega_0 & i\alpha & 0 & 0 & 0\\ 
-i\alpha & \frac{3}{2}\hbar\omega_0 & i\alpha\sqrt{2} & 0 & 0\\ 
0 & -i\alpha\sqrt{2} & \frac{5}{2}\hbar\omega_0 & i\alpha\sqrt{3} & 0\\ 
0 & 0 & -i\alpha\sqrt{3} & \frac{7}{2}\hbar\omega_0  & \cdots\\ 
0 & 0 &  0  & \vdots  & \ddots
\end{pmatrix}
\end{align}
where
\beq
   \alpha=e\frac{\phi}{L}c\sqrt{\frac{\hbar\omega_0}{2\mu c^2}}\cos(\theta).
\eeq

Assuming the system has access only to the first two levels of energy, the trunkation
\beq
\mathcal{H}=
\begin{pmatrix}
\frac{1}{2}\hbar\omega_0 & i\alpha \\ 
-i\alpha & \frac{3}{2}\hbar\omega_0 
\end{pmatrix}
\eeq
leads to the eigenvalues
\begin{equation}\label{eq:lambda}
   E_\pm=\hbar\omega_0\left[1\pm\sqrt{\frac{1}{4}+\left(\frac{\alpha}{\hbar\omega_0}\right)^2}\right].
\end{equation}

Note that when $\alpha=0$ the original energy levels $\tfrac{1}{2}\hbar\omega_0$ and $\tfrac{3}{2}\hbar\omega_0$ are recovered.

The Aharonov-Bohm effect can be veryfied by observing an enlargement of the HCl absorption lines when the coil is turned on. Particularly, the first absorption, originally a sharp line located at $\tfrac{\Delta E}{\hbar}=\omega_0$, would range from
\begin{equation}
\frac{\Delta E}{\hbar}=\omega_0,
\end{equation}
corresponding to light absorbed by molecules orthogonals to ${\bf A}$, up to
\begin{equation}
\frac{\Delta E}{\hbar}=  2\omega_0\sqrt{\frac{1}{4}+\left(\frac{e\phi c}{L\sqrt{2\hbar\omega_0\mu c^2}}\right)^2},
\end{equation}
corresponding to light absorbed by molecules parallels to ${\bf A}$.

As a final remark, we explicitly quote the expression of $\bf A$ along the $z$-axis of the torus \cite{toro}, in the Coulomb gauge and approximation of inner radius ($a$) much smaller than revolution radius ($b$),
\begin{equation}
  {\bf A}\cdot{\bf\hat z} = \frac{\mu_0}{4\pi}\frac{\pi a^2bI}{(b^2+z^2)^\frac{3}{2}}.
\end{equation}

For a torus with $10^3$ wire loops and dimensions $a=2$ cm, $b=6$ cm and for infrared light of $\hbar\omega_0\sim0.05$ eV, a current of order $\sim0.2$ A is necessary in order to
\begin{equation}
  \frac{ec{\bf A}\cdot{\bf\hat z}}{\sqrt{2\hbar\omega_0\mu c^2}} \sim 1,
\end{equation}
thus making the enlargement of the absorption lines observable.


\end{multicols}


\begin{thebibliography}{1}



   \bibitem{ab} Y. Aharonov and D. Bohm,
   Phys. Rev. {\bf 115}, 3 (1959).

   \bibitem{ab-scat} R. G. Chambers,
   Phys. Rev. Lett. {\bf 5}, 3 (1960).
   
   \bibitem{ab-toro} Akira Tonomura et. al,
   Phys. Rev. A {\bf 34}, 2 (1986).

   \bibitem{ab-var} B.R. Holstein,
   Am. J. Phys {\bf 59}, 12 (1991).   
   
   \bibitem{peshkin} M. Peshkin,
   Phys. Rev. A {\bf 23}, 1 (1981).   
   
   \bibitem{wod} K. Wódkiewicz,
   Phys. Rev. A {\bf 29}, 3 (1984).

   \bibitem{ab-bound-exp} B. D. Deaver Jr. and B. G. Donaldson,
   Phys. Lett. A {\bf 89}, 4 (1982).

   \bibitem{ring1} R.A. Webb, S. Washburn, C.P. Umbach and R.B. Laibowitz,
   Phys. Rev. Lett. {\bf 54}, 2696 (1985).      
   
   \bibitem{ring2} A.D. Stone and Y. Imry,
   Phys. Rev. Lett. {\bf 56}, 189 (1986).

   \bibitem{ring3} Alexander van Oudenaarden et. al.,
   Nature {\bf 391} (1998).
  
   \bibitem{ab-simply} S.M. Roy and V. Singh,
   J. Phys. A: Math. Gen. {\bf 22} (1989).
  
   \bibitem{toro} N.J. Carron,
   Am. J. Phys. {\bf 63}, 8 (1995).
   
   
   


\end{thebibliography}
\end{document}